\documentclass[10pt,letterpaper]{article}
\usepackage{opex3}
\usepackage{graphicx,subfigure,epsfig,epstopdf}

\def\braket#1{\mathinner{\langle{#1}\rangle}}

\newcommand{\Dp}[2]{ \frac{\partial {#1}}{\partial {#2}} }

\newcommand{\fr}[2]{\frac{{#1}}{{#2}}}

\begin{document}


\title{General Recipe for Designing Photonic Crystal Cavities} 


\author{}
\author{Dirk Englund}

\address{Department of Applied Physics, Stanford University, Stanford, CA 94305}
\author{Ilya Fushman}
\address{Department of Biophysics, Stanford University, Stanford, CA 94305}
\author{Jelena Vu\v{c}kovi\'{c}}
\address{Department of Electrical Engineering, Stanford University, Stanford, CA 94305}
\address{All authors contributed equally.}





\begin{abstract}
We describe a general recipe for designing high-quality factor (Q) photonic crystal cavities with small mode volumes. We first derive a simple expression for out-of-plane losses in terms of the $k$-space distribution of the cavity mode. Using this, we select a field that will result in a high $Q$. We then derive an analytical relation between the cavity field and the dielectric constant along a high symmetry direction, and use it to confine our desired mode. By employing this inverse problem approach, we are able to design photonic crystal cavities with $Q> 4\cdot 10^{6}$ and mode volumes V $\sim (\lambda/n)^{3}$.  Our approach completely eliminates parameter space searches in photonic crystal cavity design, and allows rapid optimization of a large range of photonic crystal cavities.  Finally, we study the limit of the out-of-plane cavity $Q$ and mode volume ratio.  
\end{abstract}



\section{Introduction}

One of the most interesting applications of photonic
crystals (PhCs) is the localization of light to very small
mode volumes -- below a cubic optical wavelength.  The principal confinement mechanism is Distributed Bragg Reflection (DBR), in contrast to, e.g., microspheres or microdisks, which rely solely on Total Internal Reflection (TIR).  However, since three-dimensional (3D) PhCs (employing DBR confinement in all 3D in space) have not been perfected yet, the mainstream of the PhC research has addressed 2D
PhCs of finite depth, employing DBR in 2D and TIR in the remaining 1D. Such structures are also more compatible with the present microfabrication and planar integration techniques. These cavities can still have small mode volumes, but the absence of full 3D confinement by DBR makes the problem of the high-quality factor (high-Q) cavity design much more challenging. The main problem is out-of-plane loss by imperfect TIR, which becomes particularly severe in the smallest volume cavities.

Over the past few years, many approaches have been proposed to address this issue, but they all focused on the optimization of a particular cavity geometry and a particular mode supported by it \cite{ref:SGJohnson2001,ref:JV2001,ref:JV2002,
Noda2003,ref:Srinivasan02,ref:Korea02a,ref:Lalanne04,ref:Ryu04,ref:Noda05,ref:Geremia02,ref:Noda2005NatureMaterials}.  Some of the proposed cavities seem promising and have already been
proven useful as components of lasers or add/drop filters.  However, a general recipe for designing optimized nanocavities is missing, and it is also not known if cavities better than the ones presently known are possible; these are exactly the issues that we will address in this article.  In Section \ref{sect2}, we will first derive a simple set of
equations for calculation of cavity $Q$ and mode volume.  In Section \ref{sect3}
we estimate the optimum $k$-space distribution of the cavity mode field.  In Section \ref{sect5}, we will finally address the question of finding PhC cavities with maximum possible figures of merit for various applications. In this process, we
start from the optimum $k$-space distribution of the cavity field; then we derive an approximate analytical relation between the cavity mode and the dielectric constant along a direction of high symmetry, and use it to create a cavity that supports the selected high-$Q$ mode in a single step. Thus, we eliminate the need for trial and error or other parameter search processes, that are typically used in PhC cavity designs. Furthermore, we study the limit of out-of-plane $Q$ factor for a given $V$ of the
cavity mode with a particular field pattern.

\section{Simplified relation between $Q$ of a cavity mode and its $k$-space Distribution}
\label{sect2}

In order to simplify PhC cavity optimization, we derive an analytical relation between the near-field pattern of the cavity
mode and its quality factor in this section.  $Q$ measures how well the cavity
confines light and is defined as

\begin{equation}
\label{eq:def_Q} Q \equiv \omega \fr{\braket{U}}{\braket{P}}
\end{equation}
where $\omega$ is the angular frequency of the confined mode.  The
mode energy is
\begin{eqnarray}
\label{eq:mode_energy} \braket{U} &=& \int \fr{1}{2} (\varepsilon
E^{2} + \mu H^{2}) dV
\end{eqnarray}
The difficulty lies in calculating $P$, the far-field radiation intensity.

Following our prior work \cite{ref:JV2001}, we consider the
in-plane and out-of-plane mode loss mechanisms in two-dimensional
photonic crystals of finite depth separately: 
\begin{equation}
\braket{P} = \braket{P_{||}}+\braket{P_{\perp}}
\end{equation}
or
\begin{equation}
\label{eq:Qbreakup}
\fr{1}{Q} = \fr{1}{Q_{||}}+\fr{1}{Q_{\perp}}
\end{equation}
In-plane confinement occurs through DBR.  For frequencies well inside the photonic band gap, this confinement can be made arbitrarily high (i.e., $Q_{||}$ arbitrarily large) by addition of PhC layers.   On the other hand, out-of-plane confinement, which dictates $Q_{\perp}$, depends on the modal k-distribution that is not confined by TIR.  This distribution is highly sensitive to the exact mode pattern and must be optimized by careful tuning of the PhC defect. Assuming that the cavity mode is well inside the photonic band gap, $Q_{\perp}$ gives the upper limit for the total $Q$-factor of the cavity mode.

\begin{figure}[htbp]
    \includegraphics[width=4in]{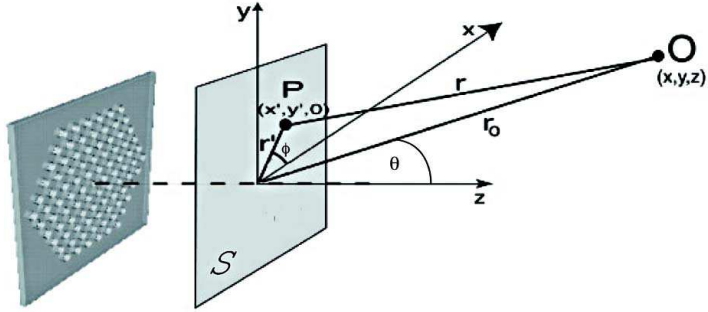}
 \caption{Estimating the radiated power and $Q_{\perp}$ from the known near field at
the surface $S$} 
\label{fig:struct-setup}
\end{figure}

Given a PhC cavity or waveguide, we can compute the near-field
using Finite Difference Time Domain (FDTD) analysis.  As described in Reference \cite{ref:JV2002}, the near-field pattern at a surface $S$ above the PhC slab contains the complete information about the out-of-plane radiation losses of the mode, and thus about $Q_{\perp}$ (Fig. \ref{fig:struct-setup}).  

The total time-averaged power radiated into the half-space above
the surface $S$ is:

\begin{equation}
P=\int\limits_0^{\pi/2} \int\limits_{0}^{2\pi} d\theta d\phi \sin(\theta) K(\theta,\phi) ,
\label{eq:prad_general}
\end{equation}

where $K(\theta,\phi)$ is the radiated power per unit solid angle.  In the appendix, we derive a very simple form for $K$ in terms of 2D Fourier Transforms (FTs) of $H_z$ and $E_z$ at the surface $S$, after expressing the angles $\theta, \phi$ in terms of $k_x$ and $k_y$:
\begin{equation}
\label{eq:K_kxky2}
K(k_x, k_y) = \fr{\eta k_z^{2}}{2
\lambda^{2}k_{\|}^{2}} \left[ \fr{1}{\eta^{2}} \left|
FT_2(E_z)\right|^{2}+
 \left| FT_2(H_z)\right|^{2} \right]
\end{equation}

Here, $\eta\equiv\sqrt{\frac{\mu_o}{\epsilon_o}}$,  $\lambda$ is the mode wavelength in air, $k=2\pi/\lambda$, and $\vec{k}_{||}=(k_x,k_y)=k(\sin\theta \cos\phi,\sin\theta\sin\phi)$ and $k_z=k \cos(\theta)$ denote the in-plane and out-of-plane $k$-components, respectively.  In Cartesian coordinates, the radiated power (\ref{eq:prad_general}) can thus be re-written as the integral over the light cone, $k_{||}<k$.  Substituting (\ref{eq:K_kxky2}) into (\ref{eq:prad_general}) gives

\begin{eqnarray}
\label{eq:Prad}
P &\approx& \fr{\eta}{2 \lambda^{2} k} \int_{k_{\|} \leq k} \fr{dk_x
dk_y}{ k_{\|}^{2}} k_z \left[ \fr{1}{\eta^{2}}  \left|
FT_2(E_z)\right|^{2}+  \left|FT_2(H_z)\right|^{2} \right]
\end{eqnarray}

This is the simplified expression we were seeking.  It gives the out-of-plane radiation loss as the light cone integral of the simple radiation term (\ref{eq:K_kxky2}), evaluated above the PhC slab.  Substituting Eq. (\ref{eq:mode_energy}) and Eq. (\ref{eq:Prad}) into Eq. (\ref{eq:def_Q}) thus yields a straightforward calculation of the $Q$ for a given mode.  In the following sections, when considering the qualitative behavior of (\ref{eq:Prad}), we will restrict ourselves to TE-like modes, described at the slab center by the triad $(E_x,E_y,H_z)$, that have $H_z$ even in at least one dimension $x$ or $y$.  For such modes, the term $|FT_2(H_z)|^2$ in (\ref{eq:Prad}) just above the slab is dominant, and  $|FT_2(E_z)|^2$ can be neglected in predicting the general trend of Q.  

The figure of merit for cavity design depends on the application: for spontaneous emission rate enhancement through the Purcell effect,
one desires maximal $Q/V$; for nonlinear optical effects $Q^2/V$; while for the strong coupling regime of cavity QED, maximizing ratios $g/\kappa \sim Q/ \sqrt{V}$ and $g/\gamma \sim 1/ \sqrt{V}$ is important. In these expressions, $V$ is the cavity mode volume: $V \equiv (\int \varepsilon(\vec{r}) |\vec{E}(\vec{r})|^{2}
d^{3}\vec{r})/\max(\varepsilon(\vec{r}) |\vec{E}(\vec{r})|^{2})$, $g$ is the emitter-cavity field coupling, and $\kappa$ and $\gamma$ are the cavity field and emitter dipole decay rates, respectively \cite{ref:JV2001}.

Thus, for a given mode pattern, we have derived a simple set of equations that allow easy evaluations of the relevant figures of merit.  In the next section, we address the problem of finding the field pattern that optimizes these figures of merit and later derive the necessary PhC to support it.

\section{Optimum $k$-space field distribution of the cavity mode field}
\label{sect3}
\begin{figure}[ht]
\begin{center}
\includegraphics[height=1.5in]{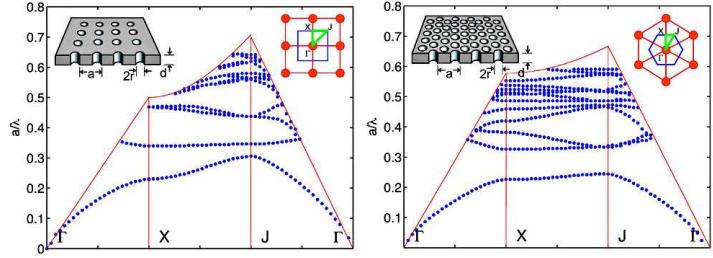}
\caption{
\textit{Left:} TE-like mode band diagram for the square lattice photonic crystal. The inset on the right shows the reciprocal lattice (orange circles), the first Brillouin zone (blue) and the irreducible Brillouin zone (green) with the high symmetry points. The inset on the left shows the square lattice PhC slab and relevant parameters: periodicity $a$, hole radius $r$, and slab thickness $d$. The parameters used in the simulation were $r/a=0.4, d/a=0.55$, and $n=3.6$.  
\textit{Right:} Band diagram for TE-like modes of the hexagonal lattice PhC. The inset on the right shows the reciprocal lattice (orange), the Brillouin zone (blue), and the first irreducible Brillouin zone (green) for the hexagonal lattice photonic crystal. The inset on the left shows the hexagonal lattice PhC slab. The parameters used for the simualtion were $r/a=0.3, d/a=0.65, n=3.6$.  A discretization of 20 points per period $a$ was used for both diagrams. }
\label{fig:PCslab}
\end{center}
\end{figure}

	The photonic crystals considered here are made by periodically modulating the refractive index of a thin semiconductor slab (see Fig. \ref{fig:PCslab}). The periodic modulation introduces an energy bandstructure for light in two dimensions $\vec{r}=(x,y)$. The forbidden energy bands are the source of DBR confinement. 

The periodic refractive index
$\epsilon(\vec{r})=\epsilon(\vec{r}+\vec{R})$ can be expanded in a Fourier sum over spatial frequency components in the periodic plane:

\begin{equation}\epsilon(\vec{r})=\sum_{\vec{G}}\epsilon_{\vec{G}}e^{i\vec{G}\cdot{\vec{r}}}\end{equation}

Here $\vec{G}$ are the reciprocal lattice vectors in the
$(k_{x},k_{y})$ plane and are defined by
$\vec{G}\cdot\vec{R}=2\pi{m}$ for integer $m$. The real and
reciprocal lattice vectors for the square and hexagonal lattices
with periodicity $a$ are:

\indent Square Lattice:
\begin{eqnarray}
\vec{R}_{mj}=ma\hat{x}+ja\hat{y}\\
\vec{G}_{ql}=\frac{2\pi{q}}{a}\hat{x}+\frac{2\pi{l}}{a}\hat{y}
\nonumber\end{eqnarray} \indent Hexagonal Lattice:
\begin{eqnarray}
\vec{R}_{mj}=ma\frac{(\hat{x}+\hat{y}\sqrt{3})}{2}+ja\frac{(\hat{x}-\hat{y}\sqrt{3})}{2}\\
\nonumber
\vec{G}_{ql}=\frac{4\pi}{a\sqrt{3}}q\frac{(-\hat{x}\sqrt{3}+\hat{y})}{2}+\frac{4\pi}{a\sqrt{3}}l\frac{(\hat{x}\sqrt{3}+\hat{y})}{2},
\nonumber\end{eqnarray}

where $m$, $j$, $q$ and $l$ are integers. The electromagnetic
field corresponding to a particular wave vector $\vec{k}$ inside
such a periodic medium can be expressed as \cite{YarivYeh}:

\begin{equation}
\vec{E}_{\vec{k}}=e^{i\vec{k}\cdot{\vec{r}}}\sum_{\vec{G}}A_{\vec{k},\vec{G}}e^{i\vec{G}\cdot{\vec{r}}}
\end{equation}
\newline

\begin{figure}[htbp]
\begin{center}
 \includegraphics[width=5.0in]{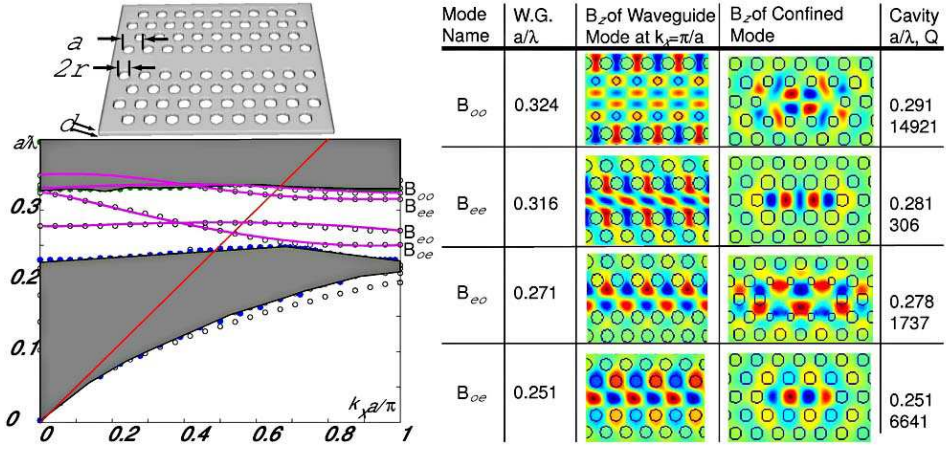}
   \caption{\textit{Left:} Band diagram for hexagonal waveguide in $\Gamma{J}$ direction, with $r/a=0.3$, $d/a=0.65$, $n=3.6$.  The bandgap (wedged between the gray regions) contains three modes.  Mode $B_{oo}$ can be pulled inside the bandgap by additional neighbor hole tuning.
\textit{Right:} $B_z$ of confined modes of hexagonal waveguide.  The modes are indexed by the $B$-field's even (``e'') or odd (``o'') parities in the $x$ and $y$ directions, respectively.  The confined cavity modes $B_{oo}$, $B_{ee}$, and $B_{eo}$ required additional structure perturbations for shifting into the bandgap.  This was done by changing the diameters of neighboring holes.  }
\end{center}
\label{fig:hex_modes_cav_wg}
\end{figure}

By introducing linear defects into a PhC lattice, waveguides can
be formed. Waveguide modes have the discrete translational
symmetry of the particular direction in the PhC lattice and can
therefore be expanded in Bloch states $E_{\vec{k}}$. In Fig.
\ref{fig:hex_modes_cav_wg}, we plot the dispersion of a waveguide in the $\Gamma{J}$ direction of a hexagonal lattice PhC.

Cavity modes can then be formed by closing a portion of a
waveguide, i.e., by introducing mirrors to confine a portion of the waveguide mode. In case of perfect mirrors, the standing wave is described by
$H=a_{k_0}H_{k_0}+a_{-k_0}H_{-k_0}$. (Here we focus on TE-like PhC
modes, and discuss primarily $H_z$, although similar relations can
be written for all other field components). Imperfect mirrors
introduce a phase shift upon reflection; moreover, the reduction
of the distance between the mirrors (shortening of the cavity)
broadens the distribution of $k$ vectors in the mode to some width
$\Delta k$. The $H_z$ field component of the cavity (i.e., a
closed waveguide) mode can then be approximated as:
\begin{equation}
\label{eq:k-space-expansion}
H_z(x,y)\sim\sum_{\vec{G}}\sum^{\vec{k}_{0}+\Delta{\vec{k}}/2}_{\vec{k}_{0}-\Delta{\vec{k}}/2}\left(A_{\vec{k},\vec{G}}e^{i\vec{k}\cdot{\vec{r}}}+A_{-\vec{k},\vec{G}}e^{-i\vec{k}\cdot{\vec{r}}}\right)e^{i\vec{G}\cdot{\vec{r}}}
\end{equation}

A similar expansion of $H_z$ can be made at the surface $S$ directly above the
PhC slab (see Fig. \ref{fig:struct-setup}), which is relevant for
calculation of radiation losses. The Fourier transform of the
above equation gives the $k$-space distribution of the cavity mode,
with coefficients $A_{\vec{k},\vec{G}}$ and
$A_{-\vec{k},\vec{G}}$. The distribution peaks are
positioned at $\pm{\vec{k_o}}\pm{\vec{G}}$, with widths directly
proportional to $\Delta k$. The $k$-space distribution is
mapped to other points in Fourier space by the reciprocal lattice
vectors $\vec{G}$. To reduce radiative losses, the mapping of
components into the light cone should be minimized
\cite{ref:JV2002}. Therefore, the center of the mode distribution
$\vec{k_{0}}$ should be positioned at the edge of the first
Brillouin zone, which is the region in $k$-space that cannot be
mapped into the light cone by any reciprocal lattice vector
$\vec{G}$ (see Fig. \ref{fig:PCslab}). For example, this region
corresponds to $\vec{k_{0}}= \pm j_x \frac{\pi}{a}\hat{k_x}\pm j_y
\frac{\pi}{a}\hat{k_y}$ for the square lattice, where $j_x, j_y
\in {0,1}$. Clearly, $|j_x|= |j_y|=1$ is a better choice for
$\vec{k_{0}}$, since it defines the edge point of the 1st
Brillouin zone which is farthest from the light cone. Thus, modes centered at this point and $k$-space broadened due to confinement, will radiate the least. Similarly, the optimum $\vec{k_0}$ for the cavities resonating in the $\Gamma J$ direction of the hexagonal lattice is
$\vec{k_{0}}= \pm \frac{\pi}{a}\hat{k_x}$ (as it is for the cavity
from Ref. \cite{Noda2003}, and for the $\Gamma X$ direction is
$\vec{k_{0}}= \pm \frac{2\pi}{a\sqrt{3}}\hat{k_y}$ (as it is for
the cavity from Refs. \cite{ref:JV2001} and \cite{ref:JV2002}).

Assuming that the optimum choice of $\vec{k}_0$ at the edge of the
first Brillouin zone has been made, the summation over $\vec{G}$
can be neglected in Eq. (\ref{eq:k-space-expansion}), because it
only gives additional Fourier components which are even further
away from the light cone and do not contribute to the
calculation of the radiation losses. In that case, we can express
the field distribution as:

\begin{equation}
\label{eq:k-space-expansion1} H_z(x,y) = \int \int dk_x dk_y A(k_x,k_y)
e^{i\vec{k}\cdot{\vec{r}}},
\end{equation}
where $A(k_x,k_y)$ is the Fourier space envelope of the mode,
which is some function centered at $\vec{k_0}=(\pm k_{0x},\pm
k_{0y})$ with the full-width half-maximum (FWHM) determined by
$\Delta\vec{k}=(\Delta k_x,\Delta k_y)$ in the $k_x$ and $k_y$
directions respectively. Eq. (\ref{eq:k-space-expansion1}) implies
that $H_z(x,y)$ and $A(k_x,k_y)$ are related by 2D Fourier
transforms. For example, if $A(k_x,k_y)$ can be approximated by a
Gaussian centered at $\vec{k_0}=(k_{0x},k_{0y})$ and with the FWHM
of $(\Delta k_x,\Delta k_y)$, the real space field distribution
$H_z(x,y)$ is a function periodic in the $x$ and $y$ directions
with the spatial frequencies of $k_{0x}$ and $k_{0y}$,
respectively, and modulated by Gaussian envelope with the widths
$\Delta x\sim 1/\Delta k_{x}$ and $\Delta y\sim 1/\Delta k_{y}$.
Therefore, the properties of the Fourier transforms imply that the
extent of the mode in the Fourier space $\Delta k $ is inversely
proportional to the mode extent in real space (i.e., the cavity
length), making the problem of $Q$ maximization even more
challenging when $V$ needs to be simultaneously minimized.
This has already been attempted in the past for a dipole cavity
\cite{ref:JV2001,ref:JV2002} and a linear defect \cite{ Noda2003, ref:Noda05, ref:Noda2005NatureMaterials},
by using extensive parameters space search. In the following
sections, we will design high $Q$ cavities by completely eliminating the need for parameter space searches and iterative trial and error approaches. 

\begin{figure}[htbp]
\begin{center}
 	\includegraphics[width=2.0in]{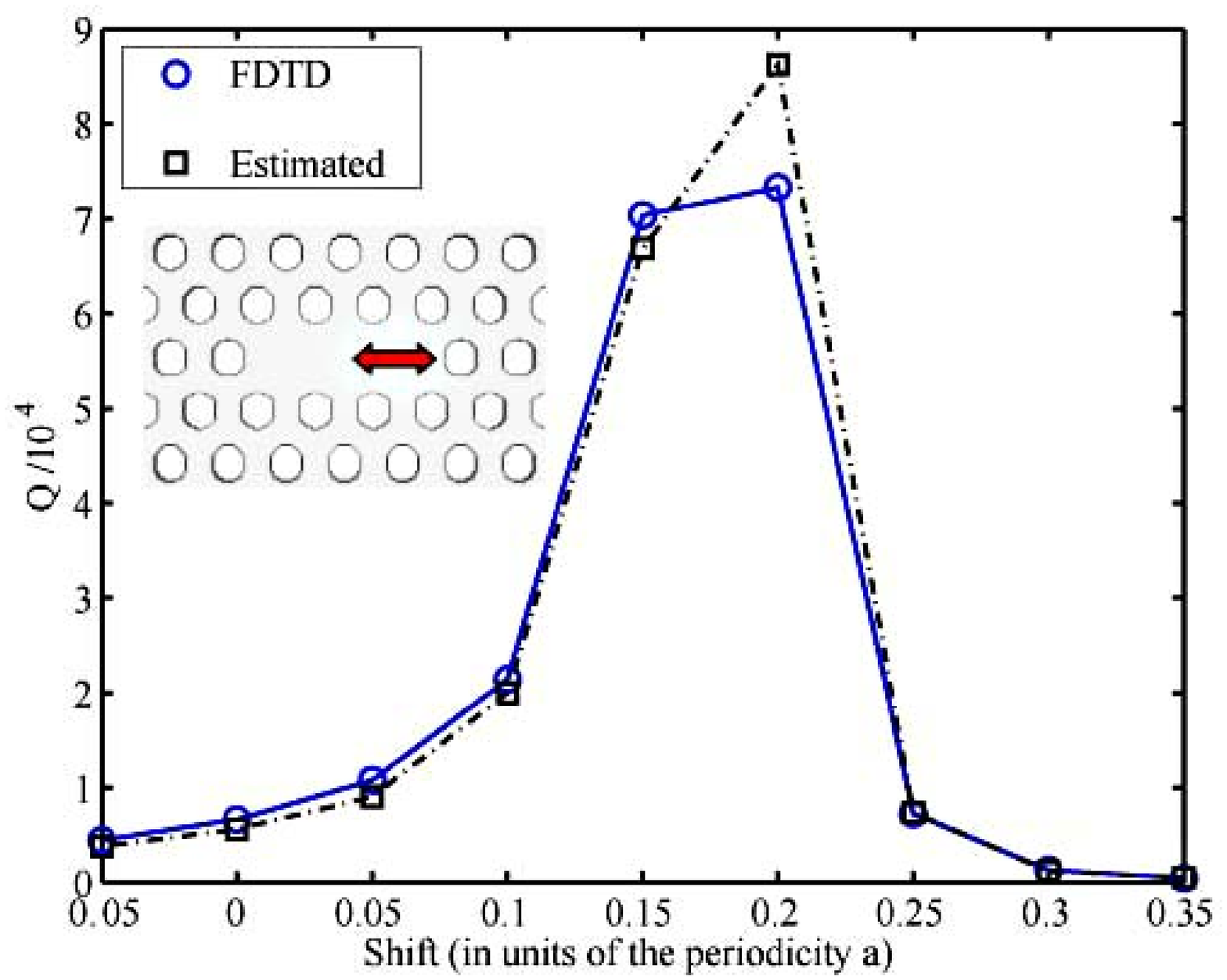}
	\includegraphics[width=2.0in]{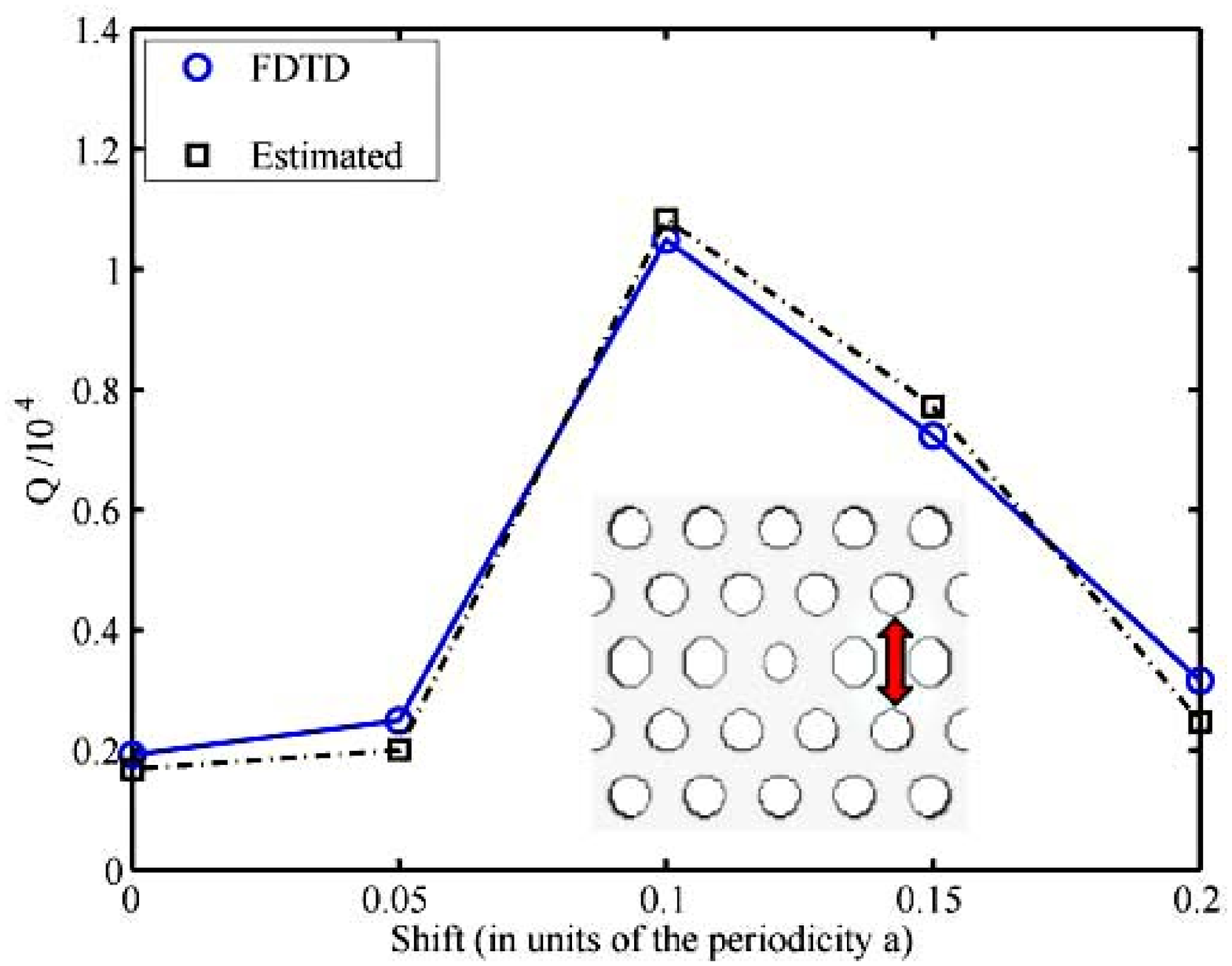}

\end {center}
    \caption{Comparison of $Q$ factors derived from Eq. (\ref{eq:Prad}) (squares) to those calculated with FDTD (circles). \textit{Left:} cavity made by removing three holes along the $\Gamma{J}$ direction confining the $B_{oe}$ mode. The $Q$ factor is tuned by shifting the holes closest to the defect as shown by the red arrow.  The $x$-axis gives the shift as a fraction of the periodicity $a$. \textit{Right:} the $X$ dipole cavity described in \cite{ref:JV2002}. The $Q$ factor is tuned by stretching the center line of holes in the $\Gamma{X}$ direction, as shown by the arrow. The $x$-axis gives the dislocation in terms of the periodicity $a$. }
    \label{fig:QV_est}

\end{figure}

	There are two main applications of Eq. (\ref{eq:Prad}). First, this formulation of the cavity $Q$ factor allows us to investigate the theoretical limits of this parameter and its relation to the mode volume of the cavity. Second, it allows us to quantify the effect of our perturbation on the optimization of $Q$ using only one or two layers of the computational field and almost negligible computational time compared to standard numerical methods. We applied Eq. (\ref{eq:Prad}) to cavities obtained from an iterative parameter space search. These cavities were previously studied in \cite{ref:JV2002} and  \cite{Noda2003}. The results for $Q$ using Eq. (\ref{eq:Prad}) at S, as well as full first-principle FDTD simulations are shown in Fig. \ref{fig:QV_est}. A good match is observed. Therefore, our expression  (\ref{eq:Prad}) is a valid measure of the radiative properties of the cavity and can be used to theoretically approach the design problem; we can also use this form to speed up the optimization of the cavity parameters. The discrepancy between Eq. (\ref{eq:Prad}) and FDTD is primarily due to discretization errors. 	

\section{Inverse problem approach to designing PhC cavities}
\label{sect5}

In the inverse approach, we begin with a desired in-plane Fourier decomposition of the resonant mode, $FT_2(\vec{H}(\vec{r}))$, chosen again to minimize radiation losses given by Eq. (\ref{eq:Prad}).  The difficulty lies with designing a structure that supports the field which is approximately equal to the target field, $\vec{H}(\vec{r})$.  

In this section, we first estimate the general behavior of $Q/V$ for structures of varying mode volume.  Then we present two approaches for analytically estimating the PhC structure $\epsilon(\vec{r})$ from the desired $k$-space distribution $FT_2(\vec{H}(\vec{r}))$.  As mentioned in Sec.\ref{sect2}, we restrict the analysis to TE-like modes $B_{eo}, B_{oe}$, and $B_{ee}$ (Fig.\ref{fig:hex_modes_cav_wg}(b)) for which we can approximate the trend of the radiation (\ref{eq:Prad}) by considering only $H_z$ at the surface $S$ just above the PhC slab.  Moreover, to make a rough estimate of the cavity dielectric constant distribution from the desired $H_z$ field on $S$, we approximate that $H_z$ at $S$ is close to $H_z$ at the slab center.
\subsection{General trend of $Q/V$}

The simplification described above allows us to study the general behavior of $Q/V$ for a cavity with varying mode volume.  Here, we assume that a structure has been found to support the desired field $H_z$.  

\begin{figure}[htbp]

\includegraphics[width=5.5in]{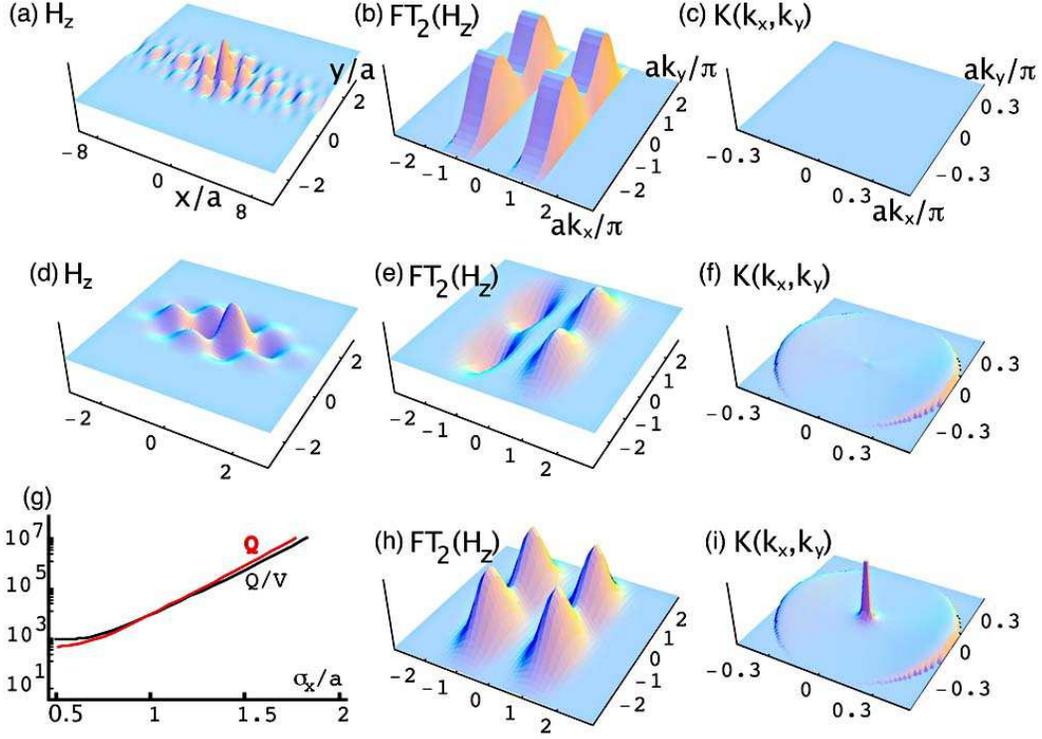}
   \caption{Idealized cavity modes at the surface S above the PhC slab; all with mode volume $\sim (\lambda/n)^{3}$. (a-c) Mode with sinc and Gaussian envelopes in $x$ and $y$, respectively: $H_z(x,y)$, $FT_2(H_z)$, and $K(k_x,k_y)$ inside the light cone; (d-g) Mode $B_{oe}$ with Gaussian envelopes in the x and y directions : $H_z(x,y)$, $FT_2(H_z)$, $K(k_x,k_y)$, and $Q_{\perp}( \sigma_{x}/a) $ as well as $Q_{\perp}/V$ . $Q_{\perp}$ was calculated using Eq. (\ref{eq:Prad}) and $E_{z}$ was neglected.; (h-i) Mode $B_{ee}$ with Gaussian envelopes in x and y can be confined to radiate preferentially upward. }

\label{fig:Ba1_calc}

\end{figure}

We again start from the expression for radiated power, Eq. (\ref{eq:Prad}), and calculate$Q$ using Eq. (\ref{eq:def_Q}).  All that is required of the cavity field is that its FT at the surface S above the slab be distributed around the four points $k_{x0}=\pm \pi/a, k_{y0}=\pm \fr{2\pi}{\sqrt{3}a}$, to minimize the components inside the light cone.  As an example, we choose a field with a Gaussian envelope in Fourier space, as described in Sec. \ref{sect3}.  For now, let us consider mode symmetry $B_{oe}$. The Fourier Transform of the $H_z$ field is then given by 
\begin{equation}
\label{eq:Gaussian_field}
FT_2(H_z)=\sum_{k_{x0},k_{y0}} sign(k_{x0}) \exp(-(k_x-k_{x0})(\sigma_x/\sqrt{2})^{2}-(k_y-k_{y0})(\sigma_y/\sqrt{2})^{2}),
\end{equation}
where $\sigma_x$ and $\sigma_y$ denote the modal width in real space.  The mode and its FT are shown in Fig.\ref{fig:Ba1_calc} (d-e).  We use Eq.\ref{eq:Prad} without the $E_z$ terms to estimate the trend in $Q$, as described above.  As the mode volume grows, the radiation inside the light cone shrinks exponentially.  This results in an exponential increase in $Q$.  This relationship is shown in Fig.\ref{fig:Ba1_calc}(g) for field $B_{oe}$ at frequency $a/\lambda=0.248$.  At the same time, the mode volume grows linearly with $\sigma_x$.  The growth of $Q/V$ is therefore dominantly exponental     , and we can find the optimal $Q$ for a particular choice of mode volume (i.e. $\sigma_x$) of the Gaussian mode cavity.

According to Fig.\ref{fig:Ba1_calc}(g), very large $Qs$ can be reached with large mode volumes and there does not seem to be an upper bound on $Q_{\perp}$. As the mode volume of the Gaussian cavity increases, the radiative Fourier components vanish exponentially, but are never zero. A complete lack of Fourier components in the light cone should result in the highest possible Q. As an example of such a field, we propose a mode with a sinc envelope in $x$ and a Gaussian one in $y$. The FT of this mode in Fig.\ref{fig:Ba1_calc}(b) is described by 
 
\begin{equation}
\label{eq:sinccos_field}
FT_2(H_z)=\sum_{k_{x0},k_{y0}} \exp(-(k_y-k_{y0})^{2}(\sigma_y/\sqrt{2})^{2}) Rect(k_x-k_{x0},\Delta k_x),
\end{equation}
where $Rect(k_x,\Delta k_x)$ is a rectangular function of width $\Delta k_x$ and centered at $k_x$.

The Fourier-transform implies that the cavity mode is described by a sinc function in $x$ whose width is inversely proportional to the width of the rectangle $Rect(k_x,\Delta k_x)$.   To our knowledge, this target field has not been previously considered in PhC cavity design. This field is shown in Fig.\ref{fig:Ba1_calc}(a-c).  Though it has no out-of-plane losses, this field drops off as $\fr{1}{r}$ and therefore requires a larger structure than the Gaussian field for confinement.  

Over the past years, many new designs with ever-higher theoretical quality factors have been suggested \cite{ref:Noda2005NatureMaterials}.  In light of our result that $Q/V$ increases exponentially with mode size, these large $Q$s are not surprising.  On the other hand, direct measurements on fabricated PhC cavities indicate that $Q$s are bounded to currently $\sim 10^{4}$ by material absorption and surface roughness \cite{imamoglu2005,englund2005}, while indirect measurements using waveguide coupling indicate values up to $6\cdot 10^{5}$\cite{ref:Noda05, ref:Noda2005NatureMaterials}.  

It is interesting to note that Eq. (\ref{eq:K_kxky2}) also allows one to calculate the field required to radiate with a desired radiation distribution.  For example, many applications require radiation with a strong vertical component; waveguide modes with even $H_z$ can be confined for this purpose so that $K(k_x,k_y)$ dominates losses at the origin in $k-$space, as shown for instance in Fig. \ref{fig:Ba1_calc}(h-i) for the confined mode pattern $B_{ee}$ of Fig.\ref{fig:hex_modes_cav_wg}.

\subsection{Estimating Photonic Crystal Design from $k$-space Field Distribution}

Now we introduce two analytical ways of estimating the dielectric structure $\epsilon(\vec{r})$ that supports a cavity field that is approximately equal to the desired field $\vec{H}_c$. These methods directly calculate the dielectric profile from the desired field distribution, without any dynamic tuning of PhC parameters, and are thus computationally fast. We focus on TE-like modes, since they see a large bandgap and exhibit electric-field maxima at the slab center. For TE-like modes, $\vec{H}_c=H_c\hat{z}$ at the center of the slab, and $\vec{H}_c \approx H_c\hat{z}$ at the surface. First, we relate $\vec{H}_c$ to one of the allowed waveguide fields $\vec{H}_w$.  The fields $H_c$ and $H_w$ at  the center of the PhC slab $(z=0)$ are solutions to the homogeneous wave equation with the corresponding refractive indices $\epsilon_c$ and $\epsilon_w$, respectively.

\begin{equation}
\label{eq:Hc}
- \mu_0 \frac{\partial^2 \vec{H_c}}{\partial{t}^2} =\omega^{2}_c \mu_{0} \vec{H}_{c} = \nabla \times \frac{1}{\epsilon_c} \nabla \times \vec{H}_c \\
\end{equation}      

\begin{equation}
\label{eq:Hw}
- \mu_0 \frac{\partial^2 \vec{H_w}}{\partial{t}^2} =\omega^{2}_w \mu_{0} \vec{H}_{w} = \nabla \times \frac{1}{\epsilon_w} \nabla \times \vec{H}_w
\end{equation}


Here $\omega_c$ and $\omega_w$ are the frequencies of the cavity and waveguide fields. We expand the cavity mode into waveguide Bloch modes:
\begin{equation} 
\label{eq:cav_mode}
H_c=\sum_{\vec{k}}c_{k}u_{k}(\vec{r})e^{i(\vec{k}\cdot\vec{r}-\omega_{k}t)} 
\end{equation}
where $u_{k}$ is the periodic part of the Bloch wave. Assuming that the cavity field is composed of the waveguide modes with $\vec{k} \approx \vec{k}_0$, we can approximate $u_k(\vec{r}) \approx u_{k0}(\vec{r})$, which leads to the slowly varying envelope approximation:
\begin{equation}
H_c \approx u_{k_0}(\vec{r}) e^{i(\vec{k_0}\cdot\vec{r}-\omega_{w} t)}\sum_{\vec{k}} c_{k} e^{i((\vec{k}-\vec{k_0})\cdot\vec{r}-(\omega_{k}-\omega_{w})t)} =  H_{w} H_{e} ,
\label{eq:cav_mode_expand}
\end{equation}
where the waveguide mode $H_{w}=u_{k_0}(\vec{r}) e^{i (\vec{k_0} \cdot \vec{r}-\omega_{w} t) } $ and the cavity field envelope $H_{e} = \sum_{\vec{k}} c_{k} e^{i((\vec{k}-\vec{k_0})\cdot\vec{r} - (\omega_{k}-\omega_{w} ) t )}$. 

The cavity and waveguide fields FT-distributions are concentrated at the edge of the Brillouin zone, where $\omega^{2}_{k} \approx \omega^{2}_{w}+\alpha\left|\vec{k}-\vec{k}_{0}\right|^2$ and $\alpha \ll 1$ (i.e., the band is nearly flat).  Differentiating (\ref{eq:cav_mode_expand}) in time twice gives: 
\begin{eqnarray}
\frac{\partial^2 H_c}{\partial{t}^2} = -\omega^{2}_{c} H_c = -\sum_{\vec{k}}c_{k}u_{k}(\vec{r})e^{i(\vec{k}\cdot\vec{r}-\omega_{k}t)} \left[\omega^{2}_{w}+\alpha\left|k-k_{0}\right|^2\right]  \approx \nonumber \\
- H_w \sum_{\vec{k}} { c_{k} e^{i((\vec{k}-\vec{k_0})\cdot\vec{r} - (\omega_{k}-\omega_{w}) t )} \left[\omega^{2}_{w}+\alpha\left|k-k_{0}\right|^2\right] } = -\omega^{2}_{w} H_w H_e+\alpha H_w \nabla^2 H_e
\label{eq:diff_cav_mode1}
\end{eqnarray}

Thus, for $\alpha\ll 1$ and finite $ \nabla^2 H_e $, $\omega_c \approx \omega_w$, i.e., the cavity resonance is very close to the frequency of the dominant waveguide mode. The condition $\alpha\ll 1$, also implies that $\omega_k \approx \omega_w$, i.e. $H_{e} \approx \sum_{\vec{k}} c_{k} e^{i(\vec{k}-\vec{k_0})\cdot\vec{r}}$.
\subsubsection{Estimating Photonic Crystal Design from $k$-space field Distribution: Approach 1}

Let us express the cavity dielectric constant $\epsilon_c$ as $\epsilon_c = \epsilon_w\epsilon_e$, where $\epsilon_e$  is the unknown envelope. $H_c$ is a solution of Eq. (\ref{eq:Hc}) and each waveguide mode $u_k(\vec{r})e^{i( \vec{k} \cdot \vec{r}-\omega_k t)}$ satisfies Eq. (\ref{eq:Hw}). From previous arguments, for $\vec{k}$ within the range corresponding to the cavity mode, $\omega_c \approx \omega_w \approx \omega_{k}$. Thus, a linear superposition of waveguide modes $\sum_{k}c_k u_k(\vec{r})e^{i( \vec{k} \cdot \vec{r}-\omega_k t)}=H_e H_w=H_c$ also satisfies Eq. (\ref{eq:Hw}), i.e. the cavity mode is also a solution of Eq. (\ref{eq:Hw}) for a slowly varying envelope. 
	We assume that the mode is TE-like, so that the $H$ field only has a $z$ component at the center of the slab. Then inserting $H_c$ from (\ref{eq:cav_mode_expand}) into Eq.(\ref{eq:Hc}) and Eq. (\ref{eq:Hw}) and subtracting the two equations with $\epsilon_c = \epsilon_w\epsilon_e$, yields a partial differential equation for $\epsilon_c(x,y)$:  
\begin{equation}
\label{eq:eps_prod_eq}
\partial_x \biggl[\fr{1}{\epsilon_w}\biggl( \fr{1}{\epsilon_e}-1\biggr) \partial_x H_c\biggr] +\partial_y\biggl[\fr{1}{\epsilon_w}\biggl( \fr{1}{\epsilon_e}-1\biggr) \partial_y H_c\biggr] \approx \mu_0 (\omega^{2}_w-\omega^{2}_c) H_c \approx 0
\end{equation}	
	
	For this approach, we consider waveguide modes with $B_{oe}$ symmetry in Fig.\ref{fig:hex_modes_cav_wg}(b).  These modes are even in $\hat{y}$, so the partial derivatives in $y$ in (\ref{eq:eps_prod_eq}) vanish at $y=0$.  The resulting simplified first-order differential equation in $1/\epsilon_e$ can then be solved directly near $y=0$, and the solution is $\fr{1}{\epsilon_w}( \fr{1}{\epsilon_e}-1) \partial_x H_c \approx C$	, where $C$ is an arbitrary constant of integration. In our analysis $\epsilon_w$ corresponds to removing holes along one line ($x$-axis) in the PhC lattice. The cavity is created by introducing holes into this waveguide, which means that $\frac{1}{\epsilon_e}-1 > 0$. The solution holds when we take the absolute value of both its sides, and for $C > 0$, this leads to the following result for the cavity dielectric constant near $y=0$:
\begin{equation}
\label{eq:eps_approach1}
\epsilon_c(x) \approx \frac{\left|\partial_{x}H_c\right|}{C+\frac{1}{\epsilon_{w}}\left|\partial_{x}H_c\right|}
\end{equation}
$C$ is a positive constant of integration, and $H_c=H_w H_e$, where $H_w$ is the known waveguide field and $H_e$ is the desired field envelope. $C$ can be chosen by fixing the value of $\epsilon_c$ at some x, leading to a particular solution for $\epsilon_c$. In our cavity designs we chose $C$ such that the value of $\epsilon_c$ is close to $\epsilon_w$ at the cavity center. To implement this design in a practical structure, we need to approximate this continuous $\epsilon_c$ by means of a binary function with low and high-index materials $\epsilon_{l}$ and  $\epsilon_{h}$, respectively.  We do this by finding, in every period $j$, the air hole radius $r_j$ that gives the same field-weighted averaged index on the $x$-axis: 
\begin{equation}
\int_{j a -a/2}^{j a + a/2} (\epsilon_h+(\epsilon_l-\epsilon_h)Rect(j a,r_j))\left|\vec{E}_c\right|^2 dx = \int_{j a -a/2}^{j a + a/2} \epsilon_c(x)\left|\vec{E}_c\right|^2 dx ,
\end{equation}
where $\vec{E}_c$ is estimated from a linear superposition of waveguide modes as $\vec{E}_c\propto \nabla\times \vec{H}_c$. We assume that the holes are centered at the positions of the unperturbed hexagonal lattice PhC holes. 

The radii $r_j$ thus give the required index profile along the $x$ symmetry axis.  The exact shape of the holes in 3D is secondary -- we choose cylindrical holes for convenience.  Furthermore, we are free to preserve the original hexagonal crystal structure of the PhC far away from the cavity where the field is vanishing.  

\begin{figure}[htbp]
\renewcommand{\baselinestretch}{1.0}
  \begin{center}
 
 \includegraphics[width=5in]{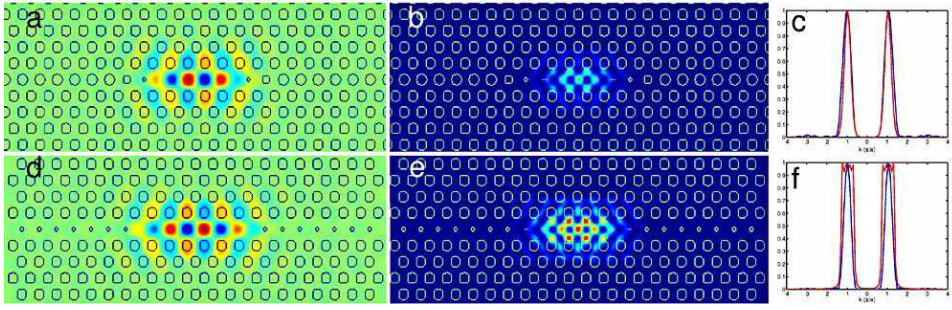}
%
    \caption{ FDTD simulations for the derived Gaussian cavity (a-c) and the derived sinc cavity (d-f).  Gaussian: (a) $B_z$; (b) $|E|$; (c) FT pattern of $B_z$ taken above the PhC slab (blue) and target pattern (red).  Sinc: (d) $B_z$; (e) $|E|$; (f) FT pattern of $B_z$ taken above the PhC slab (blue) and target pattern (red) (The target FT for the sinc cavity appears jagged due to sampling, since the function was expressed with the resolution of the simulations). The cavities were simulated with a discretization of 20 points per period a, PhC slab hole radius $r = 0.3 a$, slab thickness of $0.6 a$ and refractive index $3.6$. Starting at the center, the defect hole radii in units of periodicity a are: $(0, 0, 0.025, 0.05, 0.075, 0.1, 0.075, 0.075, 0.1, 0.125, 0.125, 0.125, 0.1, 0.125, 0.15, 0.3, 0.3)$ for the sinc cavity, and $(0.025, 0.025, 0.05, 0.1, 0.225)$ for the Gaussian cavity.}
    \label{fig:approach1}
  \end{center}
  \renewcommand{\baselinestretch}{2.0}
\end{figure}

To illustrate the power of this inverse approach, we now design PhC cavities that support the Gaussian and sinc-type modes of Eq. (\ref{eq:Gaussian_field}), (\ref{eq:sinccos_field}).  In each case, we start with the waveguide field $B_{oe}$ of Fig.\ref{fig:hex_modes_cav_wg}(b) confined in a line-defect of a hexagonal PhC.  The calculated dielectric structures and FDTD simulated fields inside them are shown in Fig. \ref{fig:approach1}.  The FT fields on S also show a close match and very little power radiated inside the light cone (Fig.\ref{fig:approach1}(c,f)).  This results in very large $Q$ values, estimated from $Q_{\perp}$ to limit computational constraints.  These estimates were done in two ways, using first principles FDTD simulations \cite{ref:JV2001}, and direct integration of lossy components by Eq. (\ref{eq:Prad}).  The results are listed in Table \ref{table:Q_derived} and show an improvement of roughly three orders of magnitude over the unmodified structure of Fig. \ref{fig:QV_est} with as small increase in mode volume. Furthermore, a fit of the resulting field pattern to a Gaussian envelope multiplied by a Sine, yielded a value of $\sigma_x/a \approx 1.6$, which, according to the plot in Fig. \ref{fig:Ba1_calc} g., puts us at the attainable limit of $Q_{\perp}$ at this mode volume.

In our FDTD simulations, we verified that $Q_{\perp}$ correctly
estimates $Q$ by noting that $Q_{\perp}$ did not change appreciably
as the number of PhC periods in the $x-$ and $y-$ directions, $N_x$
and $N_y$, was increased:  for the Gaussian-type (sinc-type) mode,
increasing the simulation size from $N_x=13,N_y=13$ ($N_x=21,N_y=9$) PhC
periods to $N_x=25,N_y=13$ ($N_x=33,N_y=13$) changed quality factors from
$Q_{||}=22\cdot 10^3,Q_{\perp}=1.4\cdot 10^6$ ($Q_{||}=17\cdot 10^3,Q_
{\perp}=4.2\cdot 10^6$) to $Q_{||}=180\cdot 10^3,Q_{\perp}=1.48\cdot
10^6$ ($Q_{||}=260\cdot 10^3,Q_{\perp}=4.0\cdot 10^6$).  (The number of
PhC periods in the x-direction in which the holes are modulated to
introduce a cavity is 9 and 29 for Gaussian and
sinc cavity, respectively, while both cavities consist of only one line
of defect holes in the y-direction.) Thus, with enough periods, the
quality factors would be limited to $Q_{\perp}$, as summarized in the
table. In the calculation of $Q$, the vertically emitted power
$\braket{P_{||}}$ was estimated from the fields a distance $\sim
0.25\cdot \lambda$ above the PhC surface.  Note that the frequencies $a/\lambda$ closely match those of the original waveguide field $B_{oe}$ ($a/\lambda_{cav}$=0.251), validating the assumption in the derivation.  

\begin{table}[htdp]
\caption{Q values of structures derived with inverse-approach 1}
\begin{center}
\begin{tabular}{c|c|c|c|c}
  & $a/\lambda_{cav}$  & $Q_{cav}$ (freq. filter) & $Q_{cav}$ (Eq. (\ref{eq:Prad})) & $V_{mode} (\frac{\lambda}{n})^3  $ \\
 \hline
Gaussian & 0.248 & $1.4\cdot 10^{6}$ & $1.6\cdot 10^{6}$ & $0.85$ \\
Sinc & 0.247 & $4.2\cdot 10^{6} $& $4.3 \cdot 10^{6} $  & $1.43$ \\
Unmodified 3-hole defect & 0.251 & $6.6\cdot 10^{3} $& $6.4\cdot 10^{3}$   & $0.63$ \\
\end{tabular}
\end{center}
\label{table:Q_derived}
\end{table}

\subsubsection{Estimating Photonic Crystal Design from $k$-space Field Distribution: Approach 2}


We now derive a closed-form expression for $\epsilon_c(x,y)$ that is valid in the whole PhC plane (instead of the center line only).  Again, begin with the cavity field $\vec{H}(\vec{r})=\hat{z}H_c$ consisting of the product of the waveguide field and a slowly varying envelope, $H_c=H_{w} H_{e},$ and treat the cavity dielectric constant as: $\fr{1}{\epsilon_c}=\fr{1}{\epsilon_{pert}}+\fr{1}{\epsilon_{w}}$.  In the PhC plane, Eq. (\ref{eq:Hc}, \ref{eq:Hw}) for a TE-like mode can be rewritten as 
\begin{eqnarray}
\label{eq:HcHw}
- \omega^{2}_c \mu_{0} H_{c} &=& \nabla \cdot (\frac{1}{\epsilon_c}\nabla H_c) \\
- \omega^{2}_w \mu_{0} H_{w} &=& \nabla \cdot (\frac{1}{\epsilon_w}\nabla H_w) 
\end{eqnarray}      
Multiplying the last equation by $H_e$, subtracting from the first, and recalling that $\omega_c\sim \omega_w$ yields
\begin{eqnarray}
\omega^{2}_w \mu_{0} H_e H_{w} -\omega^{2}_c \mu_{0}H_{c}  &=&\mu_{0}H_{c} (\omega_w^{2}-\omega_c^{2}) \approx 0\\ \nonumber
&=&\nabla \cdot (\frac{1}{\epsilon_c} \nabla H_c)-H_e  \nabla \cdot (\frac{1}{\epsilon_w} \nabla H_w)  \\
&\approx&\nabla \cdot (\fr{1}{\epsilon_{pert}} \nabla H_c)
\end{eqnarray}
where the last line results after some algebra and dropping spatial derivatives of the slowly varying envelope $H_e$.  This relation is a quasilinear partial differential equation in $1/\epsilon_{pert}$.  With boundary conditions that can be estimated from the original waveguide field, this equation can in principle be solved for $\epsilon_c $ (e.g., \cite{haberman1987}).  

Alternatively, one can find a formal solution for $\epsilon_{pert}$ by assuming a vector function $\vec{\xi}(\vec{r})$ chosen to satisfy the boundary conditions, so that 

\begin{equation}
 \fr{1}{\epsilon_{pert}} \nabla H_c = \nabla\times \vec{\xi}
 \end{equation}
 or
\begin{equation}
 \fr{1}{\epsilon_{pert}} = \fr{\nabla\times \vec{\xi} \cdot \nabla H_c^{*}}{|\nabla H_c|^{2}}
 \end{equation}
 This gives is the formal solution of the full dielectric constant $\epsilon_c=(\epsilon_{pert}^{-1}+\epsilon_{w}^{-1})^{-1}$ in the plane of the photonic crystal.  

\section{Conclusions}
We have described a simple recipe for designing two-dimensional photonic crystal cavities.  Although the approach is general, we have demonstrated its utility on the design of cavities with very large $Q> 10^{6}$ and near-minimal mode volume $\sim (\lambda/n)^{3}$. These values follow our theoretically estimated value of $Q_{\perp}/V$ for the cavity with the Gaussian field envelope, which means that we were able to find the maximal Q for the given mode volume V under our assumptions. Our approach is analytical, and the results are obtained within a single computational step. We first derive a simple expression of the modal out-of-plane radiative loss and demonstrate its utility by the straightforward calculation of $Q$ factors on several cavity designs.  Based on this radiation expression, the recipe begins with choosing the FT mode pattern that gives the desired radiation losses.  For high-$Q$ cavities with minimal radiative loss inside the light cone, we show that the transform of the mode should be centered at the extremes of the Brillouin Zone, as far removed from the light cone as possible.  Next we proved that for a cavity mode with a Gaussian envelope, $Q/V$ grows exponentially with mode volume $V$, while the cavity with the sinc field envelope should lead to even higher Q's by completely eliminating the Fourier components in the light cone. Finally, we derived approximate solutions to the inverse problem of designing a cavity that supports a desired cavity mode.  This approach yields very simple design guides that lead to very large $Q/V$.  Since it eliminates the need for lengthy trial-and-error optimization, our recipe enables rapid and efficient design of a wide range of PhC cavities.   

\subsection{Acknowledgements}
This work has been supported by the MURI Center for photonic quantum
information systems (ARO/ARDA Program DAAD19-03-1-0199). Dirk Englund
has also been supported by the NDSEG fellowship, and Ilya Fushman by the
NIH training grant

\appendix
\section{Derivation of Cavity Radiative Loss}
The radiated power per unit solid angle $K(\theta,\phi)$ can be
expressed in terms of the radiation vectors $\vec{N}$ and
$\vec{L}$ in spherical polar coordinates $(r,\theta,\phi)$ :

\begin{equation}
K(\theta,\phi)=\frac{\eta}{8\lambda^2}\biggl( \biggl |
N_\theta+\frac{L_\phi}{\eta}\biggr |^2 + \biggl |
N_\phi-\frac{L_\theta}{\eta}\biggr |^2 \biggr),
\label{eq:K_thetaphi}
\end{equation}
where $\eta=\sqrt{\frac{\mu_o}{\epsilon_o}}$. The radiation
vectors in spherical polar coordinates can be expressed from their
components in Cartesian coordinates:

\begin{eqnarray}
N_\theta&=&(N_x \cos\phi +N_y \sin \phi) \cos\theta\\ \nonumber
N_\phi&=&-N_x\sin\phi+N_y \cos \phi,
\label{eq:rad_terms}
\end{eqnarray}
and similarly for $L_\theta$ and $L_\phi$. As described in
Reference \cite{ref:JV2002}, the radiation vectors in Cartesian
coordinates are proportional to 2D Fourier transforms of the
parallel ($x$ and $y$) field components at the surface $S$ (Fig. \ref{fig:struct-setup}):

\begin{eqnarray}
N_x&=&-FT_2(H_y)\biggr |_{\vec{k}_{||}}\\ \nonumber
N_y&=&FT_2(H_x)\biggr |_{\vec{k}_{||}}\\ \nonumber
L_x&=&FT_2(E_y)\biggr |_{\vec{k}_{||}}\\ \nonumber
L_y&=&-FT_2(E_x)\biggr |_{\vec{k}_{||}}\\ \nonumber
\vec{k}_{||}&=&k(\frac{x}{r_0},\frac{y}{r_0})=k\sin\theta(\cos\phi\hat{x}+\sin\phi\hat{y}) \\ \nonumber
k_z &=& k \cos\theta,
\label{eq:rads_FTs}
\end{eqnarray}
where $k=2\pi/\lambda$, $\lambda$ is the mode wavelength in air, and $k_{||}=k\sin\theta$.

Here the 2D Fourier Transform of the function $f(x,y)$ is

\begin{eqnarray}
FT_2(f(x,y))&=&\int\int d_x d_y f(x,y) e^{i \vec{k}_{||}\cdot (x,y)} \\
             &=&\int\int d_x d_y f(x,y) e^{i (k_x x + k_y y)}
\end{eqnarray}

Substitution of expressions (\ref{eq:rad_terms}) and (\ref{eq:rads_FTs}) into (\ref{eq:K_thetaphi}) now yields an expression for the radiated power (\ref{eq:prad_general}) in terms of the FTs of the four scalars $H_x, H_y, E_x$, and $E_y$.  This expression is in general difficult to track analytically.  We will now use Maxwell's relations to express (\ref{eq:K_thetaphi}) in terms of only two scalars, $H_z$ and $E_z$.

Noting that for a bounded function $g$, $FT_2(\Dp{g}{x})=-i
k_x FT_2(g)$ (similarly for $\Dp{g}{y}$), we can re-write $N_{\theta}$ as
\begin{eqnarray}
N_{\theta} &=& \fr{k_z}{k_{\|} k} ( -k_x FT_2(H_y) + k_y
FT_2(H_x)) \\  \nonumber &=& -i \fr{k_z}{k_{\|} k}
FT_2(\Dp{H_y}{x}-\Dp{H_x}{y}) \\ \nonumber &=& \fr{k_z c
\epsilon_0}{k_{\|}} FT_2(E_z)
\end{eqnarray}
where the last step follows from Maxwell's Eq. $\vec{\nabla}\times
\vec{H} = \epsilon_o \Dp{\vec{E}}{t} = i \omega \epsilon_o \vec{E}
$. Similarly, we find $L_{\theta}=\fr{k_z c \mu_0}{k_{\|}}
FT_2(H_z)$.  From $\vec{\nabla}\cdot \vec{H} = 0$ and
$\vec{\nabla}\cdot \vec{E} = 0$ at the surface $S$, it also
follows that $N_{\phi}=\fr{-i}{k_{\|}} FT_2(\Dp{H_z}{z})$ and
$L_{\phi}=\fr{i}{k_{\|}} FT_2(\Dp{E_z}{z})$.  Substituting these
expressions into Eq. (\ref{eq:K_thetaphi}) yields
\begin{equation}
\label{K_kxky1}
K(k_x, k_y) = \fr{\eta}{8 \lambda^{2} k_{\|}^{2}} \left[
\fr{1}{\eta^{2} } \left| k_z FT_2(E_z) + i
FT_2(\Dp{E_z}{z})\right|^{2}+ \left| k_z
FT_2(H_z) + i FT_2(\Dp{H_z}{z})\right|^{2} \right]
\end{equation}

Furthermore, $E_z, H_z \propto \exp(i k_z z)$ for propagating
waves inside the light cone (which are the only ones that
determine $P$), implying that $FT_2(\Dp{E_z}{z})=-i k_z FT_2(E_z)$
and similarly for $H_z$. This allows further simplification
of the previous expression to
\begin{equation}
\label{eq:K_kxky2appendix}
K(k_x, k_y) = \fr{\eta k_z^{2}}{2
\lambda^{2}k_{\|}^{2}} \left[ \fr{1}{\eta^{2}} \left|
FT_2(E_z)\right|^{2}+
 \left| FT_2(H_z)\right|^{2} \right]
\end{equation}

Substituting this result back into the expression for total radiated power (\ref{eq:prad_general}) gives the required result (\ref{eq:Prad}):

\begin{eqnarray}
P&=&\int\limits_0^{\pi/2} \int\limits_{0}^{2\pi} d\theta d\phi \sin(\theta) K(\theta,\phi) \\ \nonumber
&=& \int_{k_{\|} \leq k}  dk_x dk_y \fr{k_{||}}{k}  K(k_x,k_y) |J(k_x,k_y)| \\ \nonumber
&=& \fr{\eta}{2 \lambda^{2} k} \int_{k_{\|} \leq k} \fr{dk_x dk_y}{ k_{\|}^{2}} k_z \left[ \fr{1}{\eta^{2}}  \left| FT_2(E_z)\right|^{2}+  \left|FT_2(H_z)\right|^{2} \right]
\end{eqnarray}

Above, $J(k_x,k_y)$ is the Jacobian resulting from the change of coordinates from $(\theta,\phi)$ to $(k_x,k_y)$.

\end{document}